\documentclass[nopreprintline,12pt,times]
{elsarticle} 
\usepackage[a4paper]{geometry}
\usepackage{etoolbox}
\makeatletter
\patchcmd{\ps@pprintTitle}
  {\def\@oddfoot}
  {\def\@oddfoot{\footnotesize\itshape
       \hfill \rm\thepage
       \hfill }}
\makeatother

\usepackage[scr=boondoxo]{mathalfa}
\DeclareFontEncoding{FML}{}{}
\DeclareFontSubstitution{FML}{fncmi}{m}{it}
\DeclareSymbolFont{fourierletters}{FML}{fncmi}{m}{it}
\SetSymbolFont{fourierletters}{normal}{FML}{fncmi}{m}{it}
\DeclareMathSymbol{\fP}{\mathalpha}{fourierletters}{`E}
\DeclareSymbolFont{Nperm}{OML}{cmss}{bx}{it}

\SetSymbolFont{Nperm}{bold}{OML}{cmss}{bx}{it}
\DeclareMathSymbol{\nw}{\mathalpha}{Nperm}{`w}

\usepackage{makecell}
\usepackage{amssymb,amsmath,yhmath,amsthm}

\usepackage{lineno}
\usepackage{ccaption}
\usepackage{tabularx}
\numberwithin{equation}{section}
\usepackage{todonotes}

\usepackage[T1]{fontenc}
\usepackage{tikz}

\usepackage{enumitem}
\usepackage{pifont}
\usepackage{epstopdf}
\usepackage{xr}
\usepackage{float}
\usepackage{hyperref}
\hypersetup{
colorlinks = true,
citecolor = blue,
linkcolor = blue
}

\PassOptionsToPackage{normalem}{ulem}
\usepackage{ulem}

  \DeclareMathAlphabet{\mathvz}{OT1}{LinuxBiolinumT-OsF}{m}{sl}

  \DeclareMathAlphabet{\mathlib}{OT1}{LinuxLibertineT-OsF}{m}{it}
  \DeclareMathAlphabet{\mathbio}{OT1}{LinuxBiolinumT-OsF}{m}{it}
\DeclareMathAlphabet{\mathpzc}{OT1}{pzc}{m}{it}

\usepackage{booktabs}

\usepackage{pdflscape}

\newcommand{\subf}[1]{\ensuremath{\left(\textsf{#1}\right)}}

\newcommand{\ag}[1]{\left[ #1 \right]}

\usepackage{algorithm,algorithmicx}
\usepackage{algpseudocode}
\usepackage[hang,flushmargin]{footmisc}

\usepackage{xspace}

\usepackage{bm}

\usepackage[draft,authormarkuptext=name,authormarkup=superscript, authormarkupposition=right]{changes}

\usepackage{tcolorbox}
\definechangesauthor[name=HKSayka, color=magenta]{1}
\definechangesauthor[name=HKSayka, color=green]{2}
\definechangesauthor[name=HKSayka, color=gray]{6}
\definechangesauthor[name=HKSayka, color=orange]{7}
\definechangesauthor[name=HKSayka, color=blue]{8}
\definechangesauthor[name=HKHK, color=black]{9}
\definechangesauthor[name=SaykaHK, color=Periwinkle]{3}
\definechangesauthor[name=SaykaSayka, color=orange]{4}
\definechangesauthor[name=HKSaykaDone, color=gray]{5}

\definecolor{Periwinkle}{rgb}{0.8, 0.8, 1.0}

\usepackage{ifthen}
\newcommand{\ShowComments}{true}
\newcommand{\forus}[1]{\ifthenelse{\equal{\ShowComments}{true}}{{\color{Periwinkle}~{#1}}}{}}

\usepackage{xfrac}

\usepackage{tabularx}
\usepackage[font={small}]{caption}

\usepackage{soul}
\setstcolor{red}

\usepackage{dutchcal}

\usepackage{afterpage}
\usepackage{todonotes}
\usepackage{setspace}

\usepackage{colortbl}
\PassOptionsToPackage{dvipsnames}{xcolor}
\usepackage{multirow}
\definecolor{maroon}{cmyk}{0,0.87,0.68,0.32}

\title{Non-classical scaling of strength with size in marine biological fibers}
\author[1]{Sayaka Kochiyama}
\author[1]{Haneesh Kesari$^*$\corref{cor1}}
\ead{haneesh\_kesari@brown.edu}
\address[1]{School of Engineering, Brown University, Providence, RI 02912, USA}

\begin{document}

\begin{abstract}

Intriguing physical phenomena observed in natural materials have inspired the development of several engineering materials with dramatically improved performance. Marine sponge glass fibers, for instance, have attracted interest in recent decades. We tested the glass fibers in tension and observed that the strength of these fibers scales inversely with their size.
While it is expected that the strength of a material scales inversely with its size, the scaling is generally believed to be inversely proportional to the square root of the specimen dimension. Interestingly, we found that the marine sponge glass fibers' strength scaled much faster, and was inversely  proportional to the square of the specimen dimension. Such non-classical scaling is consistent with the experimental measurements and classical linear elastic fracture mechanics. We hypothesize that this enhanced scaling is due to the flaw size decreasing faster than the size of the specimen. The tensile strength, as a result of non-classical, higher-order scaling, reached a value as large as  1.5 GPa for the smallest diameter specimen. The manufacturing processes through which the spicules are made might hold important lesson for further enhancing the strength of engineering materials.


\end{abstract}

\maketitle

\vspace{2pt}

\newpage
\section{Introduction}
\label{sec:Introduction}

\begin{figure}[!t]
    \centering
    \graphicspath{figure/}

        \includegraphics[width=0.7\textwidth]{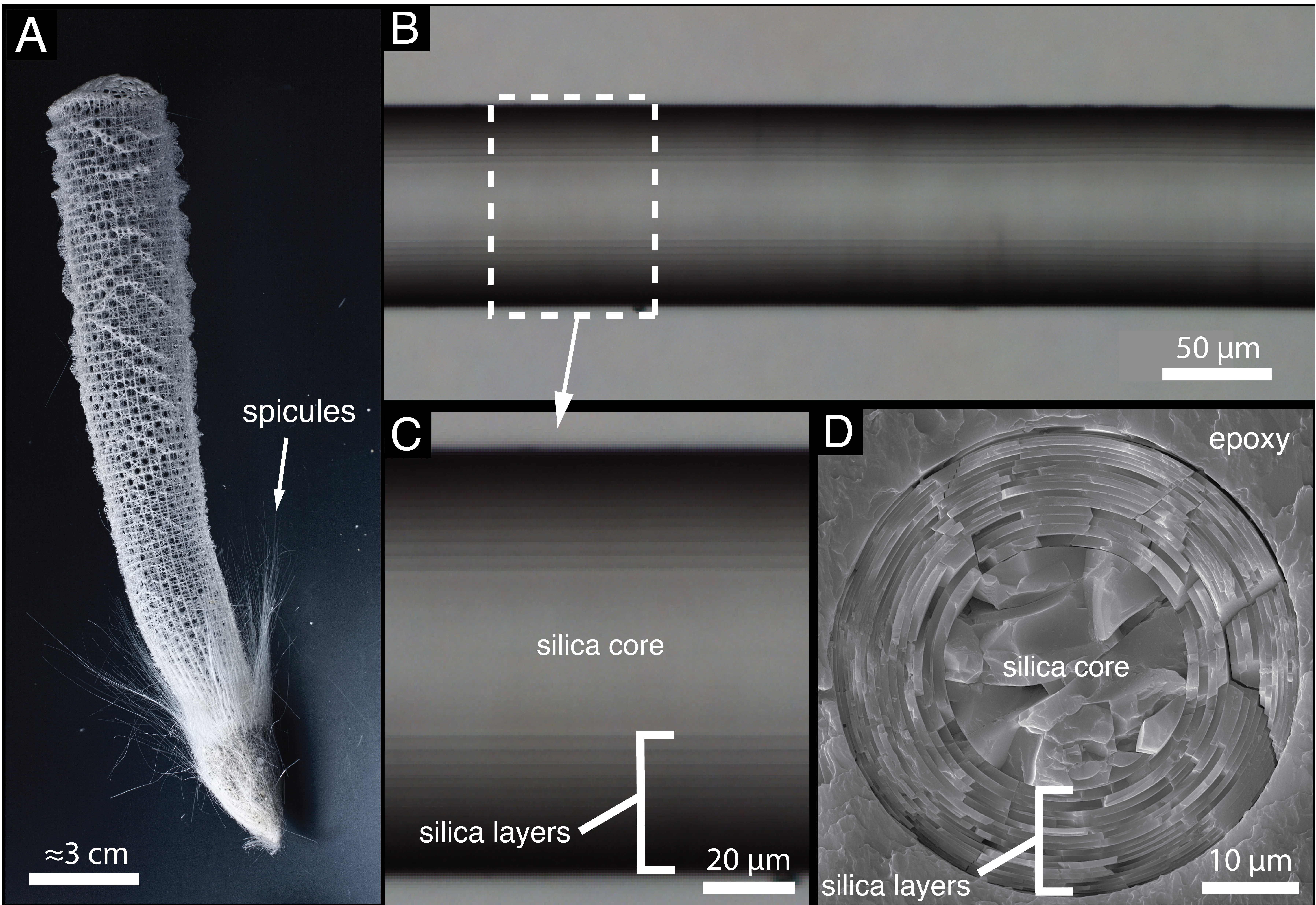}
    \caption{ \textit{Euplectella aspergillum}~sponge and its basalia spicules. 
$\subf{A}$ The entire skeletal structure of an \textit{Ea.}~sponge; the basalia spicules are identified with a white arrow (modified from~\cite{monn2015new}, with permission from National Academy of Sciences). 
$\subf{B}$ Lateral view of an \textit{Ea.}~basalia spicule taken with an optical microscope.
$\subf{C}$ A close-up view revealing the lamellar structure within the spicule.
$\subf{D}$ Cross-section of an \textit{Ea.}~basalia spicule showing its lamellar structure, imaged using a scanning electron microscope (SEM). The image clearly shows the spicule's internal architecture, consisting of a cylindrical core surrounded by coaxial layers (adapted from~\cite{monn2015new}, with permission from National Academy of Sciences).
} 
\label{fig:intro}
\end{figure}

Materials and structures found in nature can often serve as a source of inspiration towards the development of new engineering materials. One of the materials in nature that has been studied with such motivation is \textit{Euplectella aspergillum} (\textit{Ea}.), a species of marine sponge. In particular, the anchor spicules  (basalia spicules) that fasten the sponge to the ocean floor have attracted interest (see Fig.~\ref{fig:intro}\subf{A}).  Each of these basalia spicules is a fiber of several centimeters in length and approximately 50 micrometers in diameter that consist almost entirely of silica~\cite{monn2015new, weaver2007hierarchical}. Curiously, these fibers do not possess a monolithic structure, but instead has a co-axial structure consisting of a silica core surrounded by roughly 25 cylindrical layers of silica, where the adjacent pair of silica layers are separated by thin organic material (see Fig.~\ref{fig:intro}\subf{B}--\subf{D})~\cite{monn2015new, weaver2007hierarchical}. 

While the benefit---and the mechanism through which---provided by such structure is a point of active discussion, it is of interest to investigate how these biogenic silica structures with internal architecture perform in comparison to what is expected from classical linear elastic fracture mechanics theories. As discussed in \S\ref{sec:LEFMPreliminaries}\footnote{This section is aimed at providing recapitulation of the Griffith theory of fracture and size effect on material's strength ; readers may skip this section depending on the their familiarity with the concept.}, in standard linear elastic materials such as those made of silica, a material's strength is predicted to increase with decreasing specimen dimension. Would this also be the case for basalia spicules, which consist of biogenic silica and possess intricate internal architecture? If so, is the order of size effect what is classically predicted? 

As per the method described in \S\ref{sec:MaterialsMethods}, we tested the basalia spicules in tension\footnote{Tensile strength of \textit{Ea.}~spicules have also been reported by~\cite{morankar2022tensile,walter2007mechanisms}.}. The observed tensile strength was as high as 1.5 GPa for the specimen with the smallest diameter, and decreased to $\approx200$ MPa range for the larger diameter specimens.  To analyze the observed size effect in detail, we estimate the size of the crack that led to each specimens' failure using the stress analysis solution recapitulated in \S\ref{sec:CrackSize}. The results are briefly summarized and presented in \S\ref{sec:Results}. In \S\ref{sec:Discussion}, we discuss the observed size effect in detail and further comment on the measured strength values in the context of those values reported for other natural materials and technical glass. We close the paper by making some concluding remarks in \S\ref{sec:Conclusion}.




\section{Fracture Mechanics Preliminaries}
The following sections recapitulate some of the classical linear elastic fracture mechanics concepts. Readers may skip this section depending on their familiarity level.
\label{sec:LEFMPreliminaries}
\subsection{Griffith theory}
\label{subsec:GriffithTheory}
The knowledge of how strong a material is, is undoubtedly a critical part of ensuring safety in structures and engineering applications. Despite its importance, its understanding only advanced forward in the last century. For a long time, the idea that a) materials fracture at some critical applied stress, and b) such critical stress value is a material property, remained prevalent due to the intuitiveness of such notions. However, in experiments, varying strength values were measured for the same material depending upon its geometry, testing condition, and also upon treatment procedures such as polishing, etc. 

The resolution to this this apparent contradiction between the prevalent notion on material's strength and the experimental evidences was only brought about by Griffith in 1920, who recognized that the criteria for fracture should be described according to the principle of minimum energy, rather it being some critical applied stress value. Building upon the 1913 work by Inglis, who analyzed the stress concentration introduced by an elliptical hole in a uniformly stressed plate, Griffith presented the idea that the crack---which can be regarded as an infinitely narrow ellipse, and is thought to exist as flaws in many materials---extended when the increase in the total energy of the system due to the surface energy of newly created crack surfaces were balanced by the reduction of strain energy due to crack extension. For a uniformly stressed plate with a crack of length $2a_c$  perpendicular to the applied stress, Griffith's theory predicts that the
the fracture stress (strength) 
$\sigma_{\rm t}^{\rm s}$ is given as

\begin{equation}
\sigma_{\rm t}^{\rm s}=\sqrt{\frac{2E\gamma}{a_c}},
\label{eq:GriffithTheory}
\end{equation}
\noindent where $E$ and $\gamma$ are the Young's modulus and the surface energy of the material. This equation essentially predicts that the tensile strength increases inversely to the the square root of crack length. To confirm this theory, Griffith conducted an experiment on glass spherical bulbs and thin round tubes where he introduced cracks of known lengths prior to measuring their fracture strength. Indeed, he observed that the fracture strength was inversely proportional to the square root of the crack size (see Fig.~\ref{fig:GriffithPlots}\subf{A}). 

\subsection{Size effect}
\label{subsec:SizeEffect}
\begin{figure}[!h]
    \centering
\includegraphics[width=\textwidth]{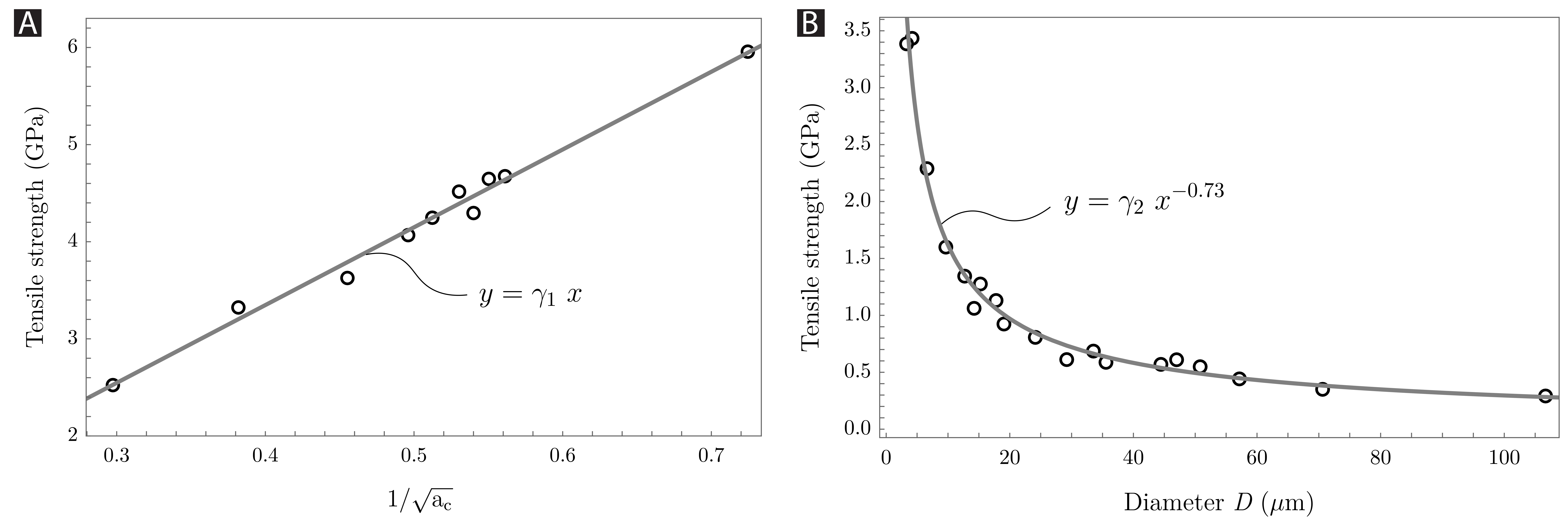}
    \caption{\subf{A} Tensile strength of glass specimens in Griffith's experiments  is plotted against $1/\sqrt{a_c}$, where $a_c$ is the length of the crack artificially introduced to the specimens and therefore is a known quantity. The measured values lie on a linear line and show that indeed, the measured tensile strength values scale as predicted by Eq.~\eqref{eq:GriffithTheory} The slope of the line of fit (gray curve) is $\gamma_1=8.28$. \subf{B} Tensile strength of the glass fibers of varying diameter $D$ tested by Griffith. A clear size effect can be observed; the nonlinear line of fit indicates that the tensile strength increased as $D^{-0.73}$, which is a stronger size effect than what the direct application of Eq.~\eqref{eq:GriffithTheory} would suggest. The scalar factor in the  nonlinear line of fit (gray curve) is $\gamma_2=8.83$. See~\cite{griffith1921vi} for original data.}
\label{fig:GriffithPlots}
\end{figure}

Griffith successfully demonstrated that the glass specimen's strength weakened with artificially introduced cracks according to his theory. However, this was not sufficient to explain why the strength values of regular materials with no visible flaws fell much short of the theoretical strength, closely related to the strength of intermolecular bond, predicted as being on the order of $E/10$. Hence, to address this question, Griffith
conducted another experiment to test the hypothesis that strength of a body is weakened due to the presence of microcracks. He reasoned that in such case, flaw size in a specimen can be inherently limited by the specimens' geometry, and hence the strength must increase with decreasing size of the specimen. He in fact reported a size effect consistent with his idea in his experiment of thin glass fibers, as can be seen in Fig.~\ref{fig:GriffithPlots}\subf{B}.

\section{Materials and Methods}
\label{sec:MaterialsMethods}
Skeletons of \textit{Ea.}~sponges were acquired from a vendor and were kept in a dry condition. The fibers in question were carefully removed from the sponge using tweezers. A several centimeter section of the smooth part of the spicule was then cut out using a razor blade to be used for the tensile test. 

The setup for the tensile test is schematically illustrated below in Fig.~\ref{fig:method}. Out of the total of 38 specimens tested, in 24 of them the two ends of the spicules were fixed by clamps to achieve a gauge length of $1.5$ cm.  In the remaining 14 samples, the gauge lengths were chosen randomly. In order to ensure that the samples did not slip with respect to the clamps as they were being tested, each end of the gauge section was marked in color using a permanent marker (slip marks, see Fig.~\ref{fig:method}); after each test, the samples were inspected to ensure that the slip marks have not moved away from the clamps, or have fractured at the clamp.
Tensile test was conducted in a load-controlled manner, where the load was applied by adding water to the loading system (Fig.~\ref{fig:method}) at the rate of 1 mL/s. Once the spicule failed, the loading system was collected and weighed. The tensile stress at failure was determined by dividing the load at failure by the cross-sectional area determined from diameter measurements, detailed in the subsequent paragraph.

\paragraph{Diameter measurement:}
For each specimen, an optical microscope was used to measure the diameter of the spicule at 6 points, where 3 points were measured close to one end and the remaining 3 points were measured close to the other end. The diameter of the spicule was determined as an average of those 6 measurements. It is worth noting that no significant tapering was observed in our spicule specimens. Upon taking the mean of the 3 diameter measurements from one end and taking the ratio to the mean of the 3 diameter measurements from the other end, we find that for the 38 spicule specimens, the mean and the standard deviation of this ratio is $1.00\pm 0.039$. Hence, we approximate our spicule specimens as having uniform cross-section along its gauge length, and use the computed diameter as a reasonable surrogate for the cross-sectional area at fracture point.

\begin{figure}[!t]
    \centering
\includegraphics[width=0.4\textwidth]{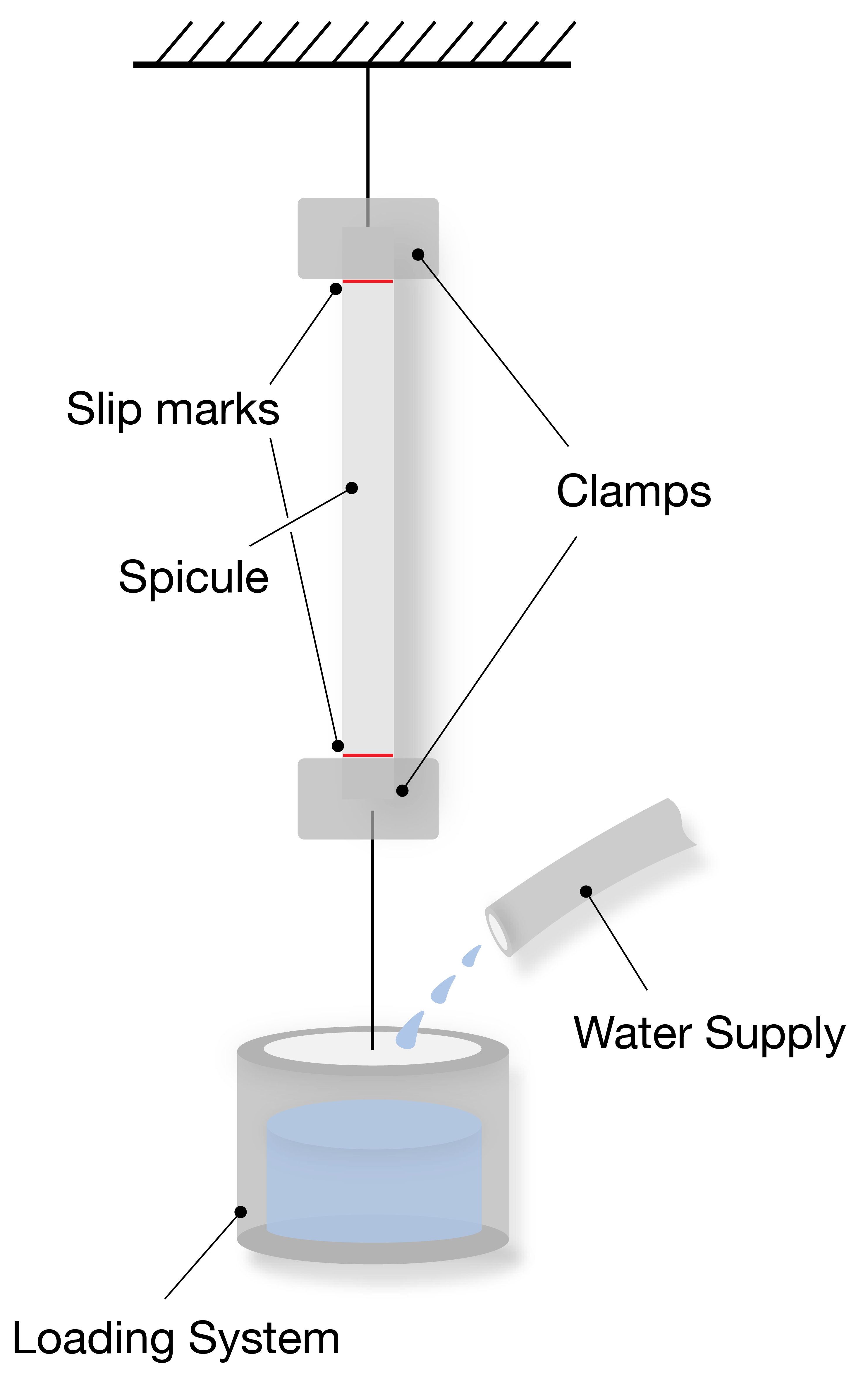}
    \caption{ Schematic of the conducted tensile test on \textit{Ea.}~basalia spicules. }
\label{fig:method}
\end{figure}

\section{Estimation of Crack Size}
\label{sec:CrackSize}
The stress intensity factor of a cylindrical structure of radius $b$ with an internal axisymmetric crack of radius $a$ that is subject to tensile load $\pi\sigma_{\rm t}^{\rm s} b^2$ has been solved for by Benthem in 1972. The result is summarized below.

\begin{subequations}
\begin{align}
K_{\rm IC}[\sigma_{\rm t}^{\rm s},a,b]:&=\frac{\sigma_{\rm t}^{\rm s} b^2}{(b^2-a^2)}\sqrt{\pi a(1-a/b)} ~f[a/b],
\end{align}
where
\begin{align}
f[a/b]:&=\frac{2}{\pi}\left(1+\frac{1}{2}\left(\frac{a}{b}\right)-\frac{5}{8}\left(\frac{a}{b}\right)^2+0.421\left(
\frac{a}{b}\right)^3\right).
\end{align}
\label{eq:KIC}
\end{subequations}

For each specimen, on substituting the measured $\sigma_{\rm t}^{\rm s}$ and $b$ values in the given definition of $K_{\rm IC}[\cdot,\cdot,\cdot]$ and equating it to a reasonable fracture toughness value for \textit{Ea.}~spicules, the internal crack size $a$ can be estimated. In Fig.~\ref{fig:result2}\subf{B}--\subf{D}, the ratio of the crack length to the specimen diameter, $a/b$, is plotted against specimen radius $b$ for three different fracture toughness values.

\section{Results}
\label{sec:Results}

\begin{figure}[!h]
    \centering
\includegraphics[width=0.6\textwidth]{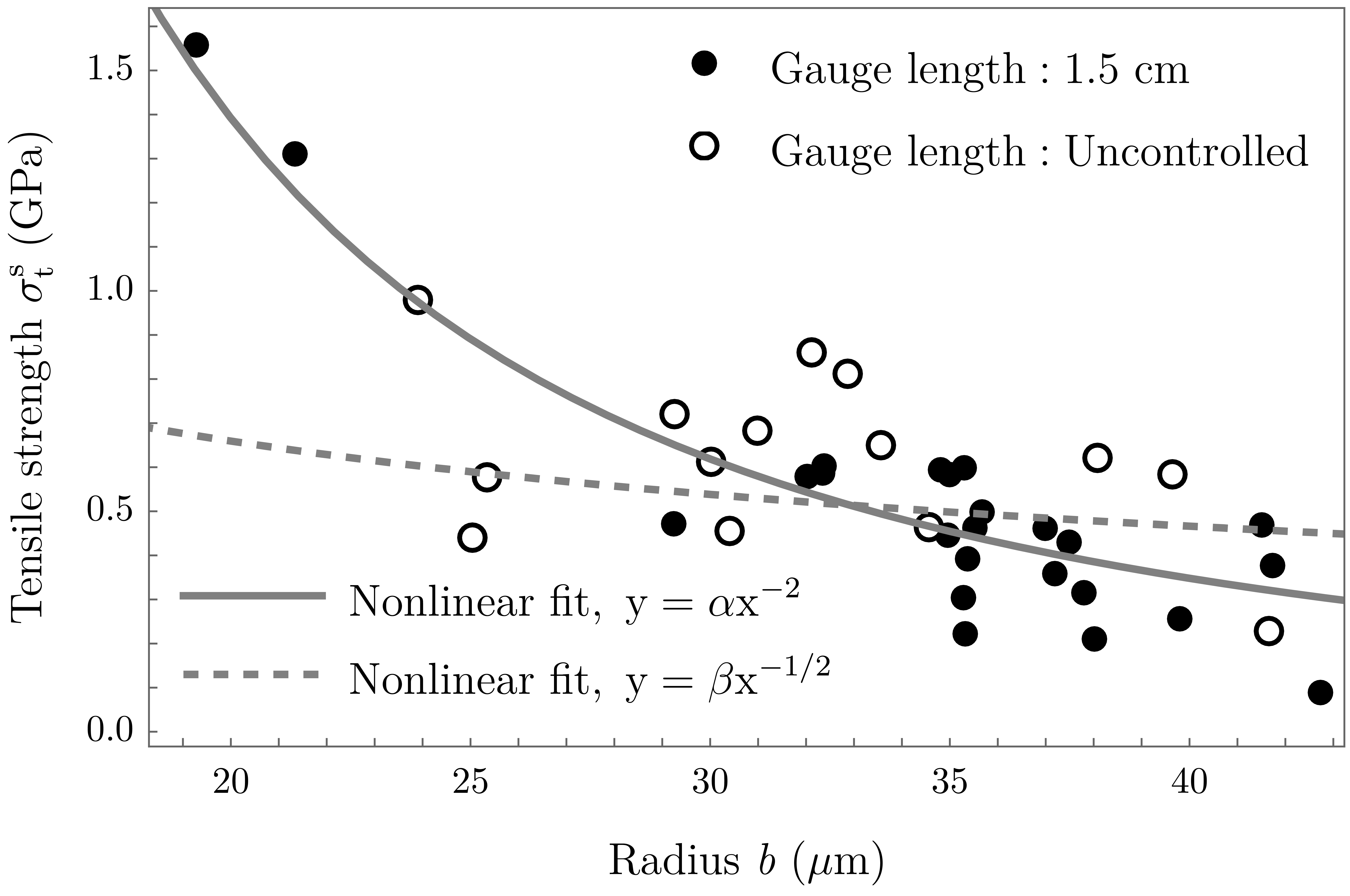}
    \caption{Tensile strength $\sigma_{\rm t}^{\rm s}$ measured for 38 \textit{Ea.}~basalia spicules. The solid curve represents the nonlinear fitting of all 38 data points, where $\alpha=557$, and the dashed curve represents the fitting of the 35 specimens whose radius $b$ is greater than $25 ~\mu$m, where $\beta=2.95$.}
\label{fig:result}
\end{figure}

\begin{figure}[!h]
    \centering
\includegraphics[width=\textwidth]{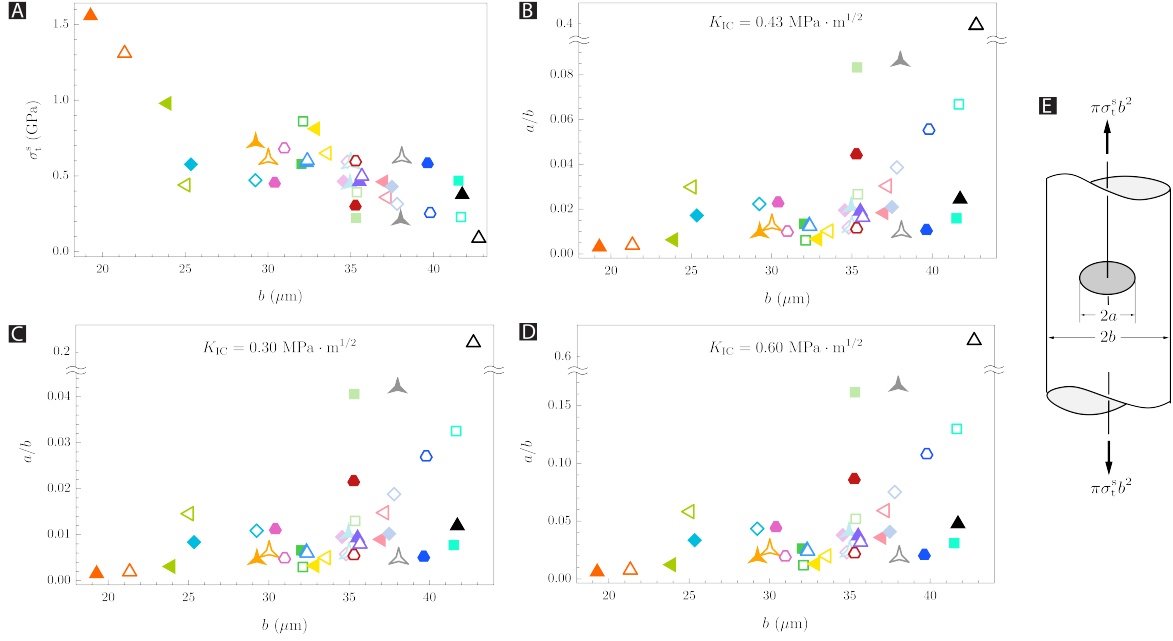}
    \caption{\subf{A} Tensile strength measured for 38 \textit{Ea.}~basalia spicules, each sample marked with a unique marker (the presented data is identical to that shown in Fig.\ref{fig:result}; the plot is presented here again to facilitate comparison with the subfigures \subf{B}--\subf{D}). \subf{B}--\subf{D} Ratio of half crack length $a$ to sample radius $b$, computed for three different fracture toughness $K_{\rm IC}$ values (0.43, 0.30, and 0.60 MPa$\cdot$ m$^{1/2}$ respectively; the value of 0.43 is provided by~\cite{morankar2022tensile}). \subf{E} Sketch of the geometry and loading of the problem considered by Benthem and for which the solution, Eq.~\eqref{eq:KIC}, was used to compute and generate the plots \subf{B}--\subf{D}. }
\label{fig:result2}
\end{figure}

The result of the tensile tests is shown in Figure~\ref{fig:result}. The tensile strength measured is $0.55\pm0.28$ GPa (mean $\pm$ standard deviation, $N=38$). When only considering the samples with the fixed gauge length of $1.5$ cm, the measured value is $0.51\pm0.32$ GPa (mean $\pm$ standard deviation, $N=24$). Within those samples whose gauge lengths were randomly chosen, the maximum and minimum gauge lengths were 5.4 cm and 1.1 cm, respectively. 

\section{Discussion}
\label{sec:Discussion}

\begin{figure}[!h]
    \centering
\includegraphics[width=0.85\textwidth]{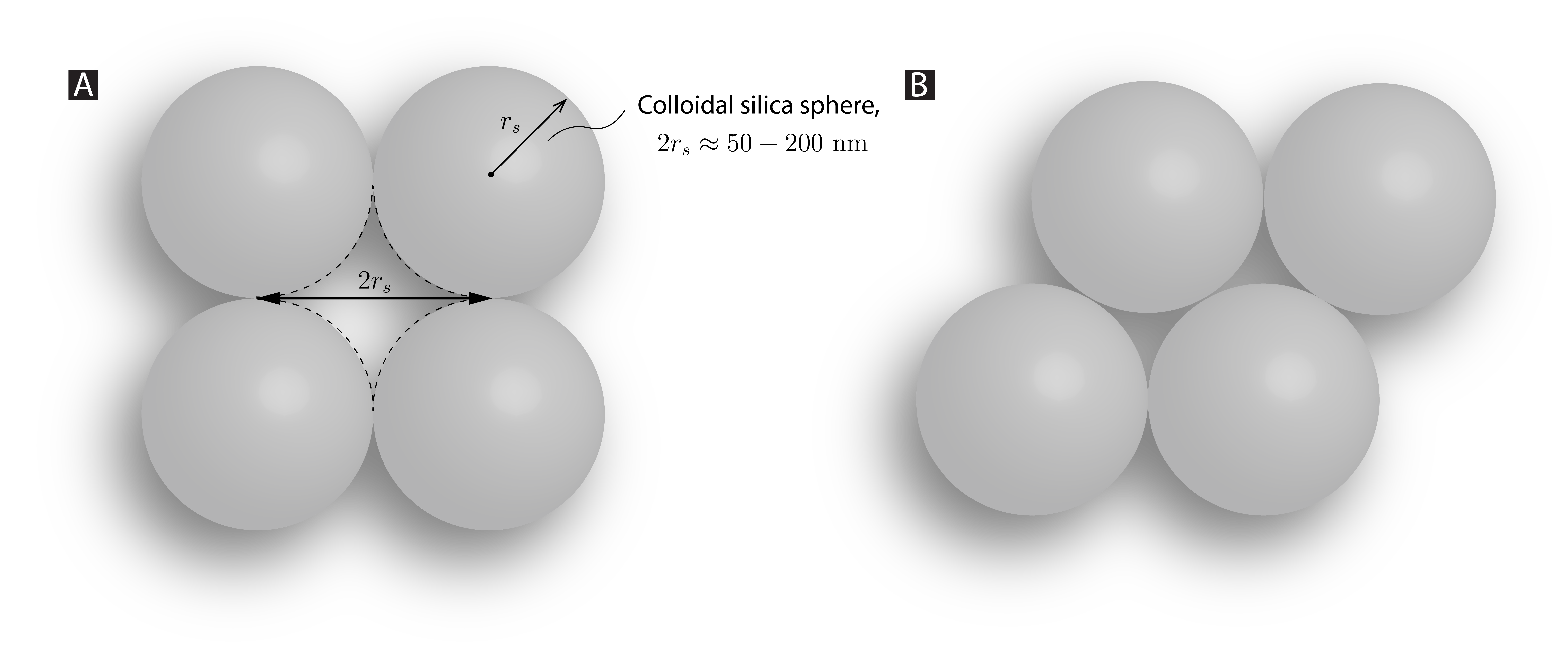}
    \caption{\subf{A} Schematic showing one way in which the colloidal silica spheres could be arranged within \textit{Ea.}~basalia spicules. When the spheres are arranged in this manner, the size of the ''gap'' (indicated by dashed enclosure) between the spheres is equal to the sphere diameter. \subf{B} The size scale of such gap is decreased as the arrangement of the spheres approach close-packing.}
\label{fig:minimumCrack}
\end{figure}

Griffith's theory given by Eq.~\eqref{eq:GriffithTheory} and the size effect are often taken together to suggest that a specimen's strength is inversely proportional to the square root of the specimens' dimension. However, in Griffith's own experiments of the glass fibers, the tensile strength increases as $D^{-0.73}$, where $D$ is the fiber diameter. The magnitude of scaling being higher than $D^{-1/2}$ suggests that the crack dimension is diminishing faster than the specimen dimension. This is certainly not an unreasonable observation; depending on the origin of the flaws/cracks formation within the specimen, it is natural to believe that ratio of the crack length to the specimen dimension can have scattered values.

Now we turn to the result shown in Fig.~\ref{fig:result}. In the figure, two nonlinear fitting lines are presented: one in the form of $y=\alpha x^{-2}$ and the other in the form of  $y=\beta x^{-1/2}$, the latter representing the classical size effect following Griffith's theory. While the data points are rather scattered, the measured $\sigma_{\rm t}^{\rm s}$ values can be well represented by the former scaling law, i.e. proportional to $b^{-2}$. While this rate of increase in the $\sigma_{\rm t}^{\rm s}$ value is much faster than what is classically expected, it is consistent with the crack length estimate shown in Fig.~\ref{fig:result2}, computed according to Eq.~\eqref{eq:KIC}.
In the figures, it can be noted that the ratio of crack diameter to specimen diameter, $a/b$, decreases with decreasing radius $b$; this is as opposed to it being a constant value over varying specimen dimension, which would be expected in the case $\sigma_{\rm t}^{\rm s}$ scaled as $b^{-1/2}$.

\begin{table}
\centering
\renewcommand{\arraystretch}{1.5} 
\setlength{\heavyrulewidth}{1.2pt} 
\setlength{\lightrulewidth}{0.8pt} 
\[
\begin{array}{l m{3cm}} 
\toprule
{\makecell{\textbf{Fiber}}} & \makecell{\textbf{Tensile Strength} \\ \textbf{(MPa)}} \\
\midrule
\rm Abaca & 430--760 \\
\rm Areca & 147--322 \\
B. mori~\rm silk & 208.45 \\
\rm Bagasse & 290 \\
\rm Bamboo & 540--630 \\
\rm Banana & 529--914 \\
\rm Coir & 175 \\
\rm Cotton & 287--597 \\
\rm Curaua & 500--1150 \\
\rm Flax & 345--1035 \\
\rm Hemp & 690 \\
\rm Henequen & 430--570 \\
\rm Isora & 500--600 \\
\rm Jute & 393--773 \\
\rm Kenaf & 930 \\
\rm Nettle & 650 \\
\rm Oil~palm & 248 \\
\rm Palf & 180--1627 \\
\rm Piassava & 134--143 \\
\rm Pineapple & 170--1627 \\
\rm Ramie & 400--938 \\
\rm Sisal & 511--635 \\
\rm Spider~silk & 875--972 \\
\rm Tussah~silk & 248.77 \\
\rm Twisted~\textit{B.~mori} ~silk & 156.27 \\
\rm Viscose & 593 \\
\rm Wood & 1000 \\
\bottomrule      
\end{array}
\]
\caption{Tensile strength of fibers found in or sourced from nature. Adapted from~\cite{gao2022structural}.}
\label{tab:StrengthTable}
\end{table}

Furthermore, it can be noted that the estimated crack size is in a physically reasonable range. Leaving out the specimen of largest diameter, the estimated crack length $2a$ lies approximately in the 120 nm--6 $\mu$m, 60 nm--3 $\mu$m, and 240 nm--12 $\mu$m range for $K_{\rm IC}$ value of $0.43, 0.30$, and $0.60$ MPa$\cdot$ m$^{1/2}$, respectively. The minimum of the range corresponds to the highest tensile strength value measured. Considering that the dimension of the colloidal silica sphere particles consisting \textit{Ea.}~spicules have been shown to be 50--200 nm in diameter, these minimum values are approaching the smallest possible feature size for any openings (see Fig.~\ref{fig:minimumCrack}). Such explanation is not inconsistent with the high tensile strength value of $1.5$ GPa measured for this specimen.

Finally, we comment on the tensile strength measured in our experiment, and how the values compare to that of other materials. Referring to Table~\ref{tab:StrengthTable}, other high strength materials known in nature---such as spider silk, bamboo fiber, and pineapple fiber---have a reported maximum value of 1--1.6 GPa. Our highest measured value of 1.5 GPa is comparable, and in fact lies among the highest measured in these materials. 

It might also interest the readers to list some of the reported values for technical glasses available in literature. One comparison could be against the measurement made by Griffith (briefly reviewed in \S\ref{sec:LEFMPreliminaries}). 
 In Fig.~\ref{fig:GriffithPlots}\subf{B}, the tensile strength of a specimen diameter of $\approx 50~\mu$m reads approximately 0.5 GPa, which is comparable with the mean value we report for basalia spicules. Study by Proctor, et al. reports a high tensile strength of nearly 6 GPa for fused silica fibers of 20-40 $\mu$m; those fibers were carefully handled to minimize any surface damage prior to the measurement so as to achieve high strength ~\cite{proctor1967strength}. Other measurements available in literature, for instance, are those of monolithic rods of $\approx300~\mu$m diameter drawn from borosilicate glass and soda-lime glass. These specimens reportedly displayed a mean tensile strength of 0.42 GPa and 0.55 GPa, respectively~\cite{stockdale1951changes}. Tensile strength values of industrial glass fibers such as E-glass and S-glass may also be mentioned here, although the silica content is only 50-60\% in composition. These glass fibers typically possess high tensile strength in the range of 2--5 GPa. 
 
The strength of spicule specimens measured in our experiments clearly fall short of some of the high strength values demonstrated by technical glasses referenced above. However, it is worthy to keep in mind that these technical glass fibers undergo controlled manufacturing/treatment process and, in some cases, careful handling process to enhance and maintain their strength.
 In such context, it is quite remarkable that a biogenic glass fiber, whose natural formation process does not involve any treatment processes (such as annealing and polishing) and was unreservedly exposed to the harsh ocean environment, can still exhibit a strength as high as 1.5 GPa.

 \section{Concluding remarks}
 \label{sec:Conclusion}
 In this paper, we presented some of the features that can be gleaned from the result of the tensile test on \textit{Ea.}~spicules. We herein summarize the key observations: 

 \begin{enumerate}
 \item The presented data shows a size effect on the spicules' tensile strength, which is inversely proportional to the square of specimen dimension.

 \item This size effect clearly deviates from the classical scaling law, which predicts inverse scaling as the square root of specimen dimension.

 \item The highest observed tensile strength value is as large as 1.5 GPa.

 \item This value is among the highest tensile strength values demonstrated by naturally sourced materials.
 \end{enumerate}

These observations render \textit{Ea.}~spicules as a potential model material.  That is to say, since the increase in strength through size effect is much more \textit{Ea.}~spicules than what has traditionally been considered possible and reasonable,  understanding of the formation process of the spicules may lead to the discovery of yet unknown strategies for achieving lighter and stronger materials.

\newpage
\bibliographystyle{elsarticle-harv}
\bibliography{RefsTensile}




%
%
%

\end{document}